\documentclass[twocolumn,aps,prl] {revtex4} 
\usepackage{graphicx}
\begin{document}
\title{Adhesion between elastic solids with randomly rough surfaces: comparison of analytical theory with molecular dynamics simulations}
\author{N. Mulakaluri and  B.N.J. Persson}
\affiliation{Institut f\"ur Festk\"operForschung, Forschungszentrum J\"ulich, D-52425 J\"ulich, Germany}

\begin{abstract}
The adhesive contact between elastic solids with randomly rough, self affine fractal surfaces is studied by molecular dynamics (MD)
simulations. The interfacial binding energy obtained from the simulations of nominally flat and curved surfaces
is compared with the predictions of the
contact mechanics theory by Persson. Theoretical and simulation results agree rather well, and most of the differences
observed can be attributed to finite size effects and to the long-range nature of the interaction between the
atoms in the block and the substrate in the MD model, as compared to the analytical theory which
is for an infinite system with interfacial contact interaction. For curved surfaces (JKR-type of problem) the effective
interfacial energy exhibit a weak hysteresis which may be due to the influence of local irreversible detachment
processes in the vicinity of the opening crack tip during pull-off.

\end{abstract}
\maketitle


Surface forces play an important role in modern technology which deals
with micro/nano scale devices. Electromechanical devices on this length scale have undergone a rapid
development within the last decade. The predominant effect of surface forces is due to the increase of
the ratio between the number of atoms on the surface and that in the volume, as the size
of an object decreases. When two surfaces are brought
together, attractive (or repulsive) forces act between them, and a non-zero force is often required to separate two solid bodies
placed in intimate contact\cite{Bowden,Johnson,BookP,Isra,JKRP,Fuller,PerssonPRL}, a phenomenon referred to as adhesion.

Adhesion manifest itself in different ways. On one hand it makes it possible for
a Gecko to walk on the ceilings or run on a vertical wall\cite{Gao,Autumn}. On the other hand, adhesion
can lead to the failure of micro or nano devices, e.g. micro-sized cantilever
beams\cite{Adhesion}. Thus, if it is too long or too thin, the free energy
minimum state corresponds to the cantilever beam (partly) bound to the substrate, which
leads to the failure of the device. However, if the surface roughness is increased,
the non-bonded cantilever state may be stabilized due to the decrease of cantilever-substrate binding energy.

Even the weakest force of interest in condensed matter physics, namely the van der Waals force,
is relative strong on a macroscopic scale. Thus for two bodies with perfectly flat surfaces in contact over $1 \ {\rm cm}^2$,
a force equivalent to the weight of a car would be necessary to separate
the surfaces even if only the weak van der Waals interaction
act between the surfaces. In practice this is not the case, and this is referred to as the {\it adhesion paradox}.
Thus the fundamental problem is not why adhesion sometimes is observed, but rather why it usually does not manifest itself
in everyday life.

The explanation for the adhesion paradox is that no surface of practical use is perfectly flat but have
surface roughness on many different length scale. In this case the interfacial bond breaking will not occur uniformly
over the contacting interface, but will start at some defect and spread by interfacial crack propagation.
The high stress concentration at the tip of cracks result in a much smaller force necessary for the separation of the surfaces than
would be expected in the ideal case of perfectly smooth surfaces and uniform bond breaking.

In a classical study, Fuller and Tabor\cite{Fuller} have studied the influence of surface roughness on adhesion. They used
silicon rubber balls in contact with (Plexiglas) surfaces with different amount of surface roughness produced by
sand blasting. They found that with increasing surface roughness, the pull-off force dropped rapidly to zero,
and using elastically softer rubber resulted in stronger adhesion than for stiffer rubber. This experimental work,
and other similar studies\cite{PerssonPRL}, have stimulated much work to understand the role of roughness on adhesion.

Many practical application of adhesion involves soft elastic solids, e.g., pressure sensitive adhesives usually consist of
weakly crosslinked rubber, which may exhibit very complex
processes during pull-off such as cavitation and stringing\cite{all,gorb}. Here we consider the
influence of surface roughness on adhesion for the most simple case of a purely elastic solid in contact with a hard randomly rough
surface. If this simplest case cannot be understood in detail there is no hope to understand much more complex
systems involving complex rheological materials in contact with randomly rough surfaces.

In order for two elastic solids with rough surfaces to make adhesive contact it is necessary to deform the
surfaces elastically, otherwise they would only make contact in three points and the adhesion would vanish,
at least if the spatial extent of the adhesion forces is neglected. Deforming the surfaces to increase the contact
area $A$ result in some interfacial bonding $-\Delta \gamma A$ (where $\Delta \gamma = \gamma_1+\gamma_2-\gamma_{12}$ is the
change in the interfacial energy per unit area upon contact), but it cost elastic deformation energy $U_{\rm el}$
which will reduce the effective binding. That is, during the removal of the block from the substrate the elastic
compression energy stored at the interface is given back and helps to break the adhesive bonds in the area
of real contact. Most macroscopic solids does not adhere with any measurable force,
which imply that the total interfacial energy $-\Delta \gamma A + U_{\rm el}$ vanish, or nearly
vanish in most cases. However, not all the stored elastic energy $U_{\rm el}$ may be used to break adhesive bonds during
pull-off but some fraction of it may be radiated as elastic waves (phonons) into the solids.
This would result in an increase in the effective interfacial binding energy, and would result in adhesion hysteresis.

In this letter we study the variation of the effective interfacial binding energy with the surface roughness amplitude,
for elastic solids with randomly rough surfaces.
We assume perfect (linear) elasticity and compare the results of MD-simulations for nominally flat and curved surfaces with
a recently developed contact mechanics theory\cite{P1,P2,P3,P4}.
The theoretical and simulation results agree rather well, and most of the differences
observed can be attributed to finite size effects and to the long-range nature of the interaction between the
atoms in the block and the substrate in the MD model, as compared to the analytical theory which
is for an infinite system with interfacial contact interaction. For curved surfaces (JKR-type of problem) the effective
interfacial energy exhibit a weak hysteresis which may be due to the influence of local irreversible detachment processes.
This study represents the first test of the theory prediction for the effective adhesion energy for 3D systems\cite{CarboneS}.

We review the contact mechanics theory of Persson briefly.
It can be used to calculate the
stress distribution at the interface, the area of real contact and the average interfacial
separation between the solid walls\cite{P1,P4}. In this theory, the interface is studied at different
magnifications $\zeta = L/\lambda$ where $L$ is the linear size of the system and $\lambda$
the resolution. The wavevectors are defined as $q=2 \pi /\lambda$ and $q_L=2 \pi /L$ so that
$\zeta=q/q_L$.

Consider an elastic block with a flat
surface in adhesive contact with a hard substrate with a randomly rough surface.
Let $\sigma({\bf x},\zeta)$ denote the (fluctuating) stress at the interface between the solids when
the system is studied at the magnification $\zeta$. The distribution of interfacial stress
$$P(\sigma,\zeta) = \langle \delta (\sigma- \sigma({\bf x},\zeta))\rangle. \eqno(1)$$
In this definition we do not include the $\delta (\sigma )$-contribution from
the non-contact area.

For perfect (or complete) contact it is easy to show that $P(\sigma,\zeta )$ satisfies\cite{P1}
$${\partial P \over \partial \zeta} = f(\zeta) {\partial^2 P\over \partial \sigma^2},\eqno(2)$$
where
$$f(\zeta ) = {\pi \over 4} {E^*}^2 q_L q^3 C(q).$$
Here $E^* = E /(1-\nu^2)$ is the effective elastic modulus.
The surface roughness power spectrum
$$C(q) = {1\over (2\pi )^2}\int d^2x  \ \langle h({\bf x})h({\bf 0})\rangle e^{-i{\bf q}\cdot {\bf x}}$$
where $z=h({\bf x})$ is the surface height at the point ${\bf x}=(x,y)$ and where $\langle .. \rangle$
stands for ensemble average. The basic idea is now to assume that (2)
holds locally also for incomplete
contact.

To solve (2) one needs boundary conditions. If we assume that,
when studying the system at the lowest magnification $\zeta=1$
(where no surface roughness can be observed, i.e.,
the surfaces appear perfectly smooth),
the stress at the interface is constant and equal to $p=F_{\rm N}/A_0$,
where $F_{\rm N}$ is the load and $A_0$ the nominal contact area, then
$P(\sigma, 1) = \delta (\sigma -p)$.
In addition to this ``initial condition'' we need two boundary conditions along
the $\sigma$-axis. Since there can be no infinitely large stress at the interface we require
$P(\sigma, \zeta) \rightarrow 0$ as $\sigma \rightarrow \infty$.
For adhesive contact, which interests us here, tensile stress occurs at the interface close
to the boundary lines of the contact regions. In this case we have the boundary condition
$P(-\sigma_{\rm a},\zeta)=0$, where $\sigma_{\rm a}>0$ is the largest tensile stress possible.
The detachment stress $\sigma_{\rm a} (\zeta)$ depends on the magnification and can be related to
the effective interfacial energy (per unit area) $\gamma_{\rm eff}(\zeta)$ using
the theory of cracks\cite{P3}
$$\sigma_{\rm a} (\zeta) \approx \left ({\gamma_{\rm eff}(\zeta) E q \over 1-\nu^2}\right )^{1/2},$$
where
$$\gamma_{\rm eff}(\zeta) A^*(\zeta) = \Delta \gamma A^*(\zeta_1)- U_{\rm el} (\zeta),\eqno(3)$$
where $A^*(\zeta)$ denotes the total contact area at the magnification $\zeta$,
which is larger than the projected contact area $A(\zeta)$. $U_{\rm el} (\zeta)$ is the elastic energy stored
at the interface due to the elastic deformation of the solids on length scale shorter than $\lambda = L/\zeta$,
necessary in order to bring the solids into adhesive contact (see below).

From (1) it follows that
the area of apparent contact (projected on the $xy$-plane)
at the magnification $\zeta$, $A(\zeta)$, normalized
by the nominal contact area $A_0$,
can be obtained from
$${A(\zeta)\over A_0} = \int_{-\sigma_{\rm a}(\zeta)}^\infty d\sigma \ P(\sigma, \zeta).\eqno(4)$$
We denote $A (\zeta) / A_0 =P_p (q)$, where the index $p$ indicates that $A (\zeta) / A_0$
depends on the applied squeezing pressure $p$.
The area of (apparent) contact at the highest magnification $\zeta=\zeta_1$ gives the real contact area.
For the elastic energy $U_{\rm el}$ we use\cite{YP}
$$U_{\rm el} \approx A_0 E^* {\pi \over 2} \int_{q_L}^{q_1} dq \ q^2W(q,p)C(q),\eqno(5)$$
where $q_L$ and $q_1$ are the smallest and the
largest surface roughness wave vectors, and\cite{YP}
$$W(q,p) = P_p(q) \left [\beta +(1-\beta) P_p^2(q)\right ],$$
where $\beta = 0.45$.
The equations given above are solved as described in Ref.~\cite{P2}.

Let us provide some details about the numerical simulations.
The molecular dynamics system has lateral dimension $L_{x}=N_{x}a$ and $L_{y}=N_{y}a$,
where $a$ is the lattice spacing of the block.
In order to accurately study contact mechanics between elastic solids,
it is necessary to consider that the thickness of the block is (at least) of the same
order of the lateral size of the longest wavelength roughness on the substrate.
We have developed a multiscale MD approach to study contact mechanics\cite{YTP}.
Periodic boundary condition has been used in $xy$ plane.
For the block $N_{x}=N_{y}=512$, while the lattice space of the substrate
$b=a$.
The mass of the block atoms is 197 a.m.u. and the $a=2.6 \ \rm \AA$.
The elastic modulus and Poisson ratio of the block are
$E=77.2 \ {\rm GPa}$ and $\nu=0.42$.
For self-affine fractal surfaces, the power spectrum has power-law behavior $C(q) \sim q^{-2(H+1)}$,
where the Hurst exponent $H$ is related to the fractal dimension $D_{\rm f}$ of the surface
via $H=3-D_{\rm f}$. For real surfaces this relation holds only for a finite wave vector region
$q_0<q<q_1$.
Note that in many cases, there is a roll-off wave vector $q_0$ below which
$C(q)$ is approximately constant.
Here $q_L=2\pi/L, q_0=3q_L, q_1=512 q_L$.
The physical meaning is that by choosing $q_0=3q_L$ one can obtain a self-average equivalent to an average over $9$ independent samples.
In MD simulations, the substrate is rigid and fractal with fractal dimension
$D_{\rm f}=2.2$ and root-mean-square roughness is varied from $h_{\rm rms}=0.1~{\rm nm}$ to $2.5~{\rm nm}$.
We also studied adhesion where the substrate was curved into a nominally spherical cup with the radius of curvature $R=4410 \ {\rm \AA}$.
The calculations are carried out at temperature $T=0 \ {\rm K}$.

\begin{figure}[h!]
\includegraphics[width=0.3\textwidth,angle=0]{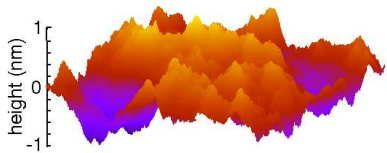}
\caption{\label{topography.3Angstrom}
The surface topography of a randomly rough surface with the root-mean-square roughness
$0.3 \ {\rm nm}$ and the Hurst exponent $H=0.8$ (i.e., fractal dimension $D_{\rm f} = 2.2$).
}
\end{figure}

\begin{figure}
\includegraphics[width=0.4\textwidth,angle=0]{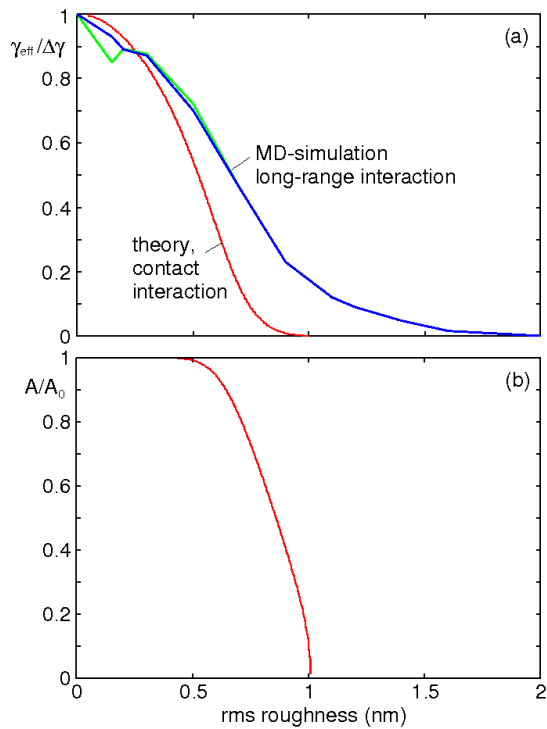}
\caption{\label{1rms.2gamma.exp.theory}
(a) The interfacial binding energy $\gamma_{\rm eff}$ (in units of the interfacial binding energy $\Delta \gamma = \gamma_1+\gamma_2-\gamma_{12}$
for flat surfaces) as a function of the root-mean-square roughness.
The blue and green curves are the result of the MD-simulations obtained during contact formation (approach) and
contact breaking (pull-off). The red curve is the theory prediction (using elastic continuum mechanics) for an
infinite system with interfacial contact interaction.
(b) The area of real contact $A$ as a function of the root-mean-square roughness calculated
using the contact mechanics theory.
The substrate surfaces are self-affine fractal (randomly rough) with the Hurst exponent $H=0.8$.
}
\end{figure}

The atoms at the interface between block and substrate interact with
the potential
$$V(r)=4\epsilon \left[ \left(\frac{r_0}{r} \right)^{12} -\left( \frac{r_0}{r} \right)^{6} \right] $$
where $r$ is the distance between the pair of atoms. The parameter
$\epsilon$ is the binding energy between two atoms at separation $r=2^{1/6}r_{0}$.
In the calculations presented below we have used the $r_{0}=3.28 \ {\rm \AA}$ and
$\epsilon=100.0 \ {\rm meV}$. In order to calculate the interfacial binding energy, first we move the (upper surface)
block at a low constant velocity towards substrate until the total
force on the bottom layer atoms of the block vanishes. At this point the work done by the substrate on the bottom layer of the block is
defined as $\gamma_{\rm eff} A_0$. In a similar way we calculate $\gamma_{\rm eff}$ during the separation
of the block from its equilibrium separation to infinite separation (pull-off).
We find, as expected, that to within the accuracy of our calculations, $\gamma_{\rm eff}$ obtained on approach and during
separation are identical. For a flat substrate $\Delta \gamma$ turns out to be $\approx 100 \ {\rm meV /\AA^2}$.

In Fig. \ref{1rms.2gamma.exp.theory}(a)
we show the interfacial binding energy $\gamma_{\rm eff}$, in units of the interfacial binding energy $\Delta \gamma = \gamma_1+\gamma_2-\gamma_{12}$
for flat surfaces,  as a function of the root-mean-square roughness.
The blue and green curves are the result of the MD-simulations obtained during contact formation (approach) and
contact breaking (pull-off). The red curve is the theory prediction (using elastic continuum mechanics) for an
infinite system with wall-wall contact interaction. The general shape of the curves are the same, but in the  MD calculation the adhesion
extend to larger surface roughness which we attribute to the small system size and the long range wall-wall interaction used
in the MD model. That is, in the MD model there is an attractive force even between the non-contact surfaces which is absent in the
continuum mechanics theory. This long-range attractive interaction is particular important for small systems and for surfaces with
small amplitude roughness.

We note that the question of interaction energy at a distance
is related to the difference between JKR\cite{JKRP} and DMT theories\cite{Maugis1}.
However, the situation is more complex in our case
because here we have surface roughness which may generate ``large'' surface areas
where the solid walls are closely spaced. In the JKR and DMT theories (involving smooth curved surfaces) the surfaces
are only closely spaced close to the rim of the (circular) contact area.
In our continuum mechanics theory we could in principle take
(approximately) into account the interaction between the non-contact surface area
using the (calculated) distribution of interfacial separations $P(u)$\cite{Carlos}.

In Fig. \ref{1rms.2gamma.exp.theory}(b)
we show the calculated area of real contact $A$, at the point where the external load vanish, as a function of the root-mean-square roughness calculated
using the contact mechanics theory. Note that, as expected, $A$ vanish for the same roughness amplitude where $\gamma_{\rm eff} = 0$.

\begin{figure}
\includegraphics[width=0.4\textwidth,angle=0]{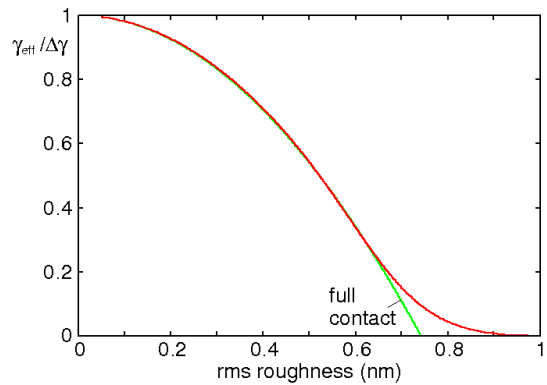}
\caption{\label{gamma.full.exact}
The interfacial binding energy $\gamma_{\rm eff}$ (in units of the interfacial binding energy $\Delta \gamma = \gamma_1+\gamma_2-\gamma_{12}$
for flat surfaces) as a function of the root-mean-square roughness.
The red line is the full theory (from Fig. \ref{1rms.2gamma.exp.theory}) while the green line is the result for the interfacial energy
assuming full contact ($A=A_0$) for all rms roughness values.
}
\end{figure}

In an earlier paper \cite{PerssonPRL} it has been shown that the pull-off force between silicon rubber balls and surfaces with
different types of (non-fractal) roughness could be explained by a very simple theory where $\gamma_{\rm eff}$
was calculated assuming that complete contact ($A=A_0$) occur between the rubber and the substrate in the nominal (or apparent) contact area.
Only for large roughness, where the pull-off force was $\sim 20\%$ or less of the pull-off for the flat substrate,
did this approach underestimate the pull-off force. This is consistent with the theory presented above. Thus in Fig. \ref{gamma.full.exact}
we show the interfacial binding energy $\gamma_{\rm eff}$ as a function of the root-mean-square roughness using the full theory
(red line, from Fig. \ref{1rms.2gamma.exp.theory}), and the result obtained assuming full contact
for all rms roughness values (green line). Only for $\gamma_{\rm eff} /\Delta \gamma < 0.2$ does the full-contact theory underestimate
the interfacial binding energy, which is consistent with the experimental data presented in Ref. \cite{PerssonPRL}.

\begin{figure}
\includegraphics[width=0.45\textwidth,angle=0]{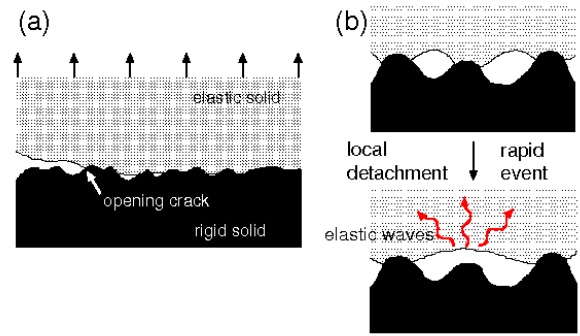}
\caption{\label{openingcrack}
(a) An opening of crack during pull-off. (b) A local detachment event in front of the crack tip. Asperity contact region detach rapidly with elastic energy radiating inside the block rather than being used to break
other asperity contact regions (schematic).
}
\end{figure}

Most solid objects have some macroscopic curvature which must be taken into account when determining the force necessary
to separate two solids in adhesive contact. The theoretically most well defined situation is
the contact between a ball and a flat surface. In this case, if at least one of the solids is
elastically soft enough, the so called JKR theory can be applied, which
predict the pull-off force to be\cite{JKRP}
$$F= {3\pi \over 2} \gamma_{\rm eff} R. \eqno(6)$$
In this equation $\gamma_{\rm eff}$ is the effective interfacial energy (usually denoted as the work of adhesion)
which includes the influence of the surface roughness on the interfacial binding energy. During pull-off the interfacial
bond-breaking will not occur everywhere simultaneously, but
a (circular) interfacial opening crack will propagate at the interface,
see Fig. \ref{openingcrack}(a). The energy to propagate an interfacial crack may be larger
than the energy necessary just to break the interfacial bonds. Thus, the work of adhesion
which enter in (6) may be larger than the effective interfacial energy given by (3).
For example, for rubber materials, unless the crack moves extremely slowly,
there may be a large energy dissipation in the rubber close to the crack tip
caused by the viscoelasticity of the rubber, which may enhance the force necessary for
pull-off by several order of magnitudes. This effect can be taken into account by multiplying the $\gamma_{\rm eff}$ given by
(3) with a factor $1+f(v,T)$, where $f(v,T)$ is caused by viscoelastic energy dissipation in front of the crack tip,
which depends on the crack tip velocity $v$ and the temperature $T$\cite{Gent,Maugis}. And even for
perfectly elastic solids the work of adhesion may be larger than given by (3) since the separation of the surfaces at the crack tip
may involve rapid events caused by local elastic instabilities close to the crack tip. Thus, for example,
if a low asperity makes contact with the rubber the rubber may
suddenly detach when the tensile stress from the approaching
crack becomes high enough, see Fig. \ref{openingcrack}(b). If the elastic energy stored in the vicinity of the contact area before
detachment is radiated as elastic waves into the block, rather than used to break other asperity contact regions,
then this process will enhance the work of adhesion $\gamma_{\rm eff}$ and the pull-off force. We will denote processes of the type illustrated in
Fig. \ref{openingcrack}(b) as local irreversible detachment processes.
This mechanism has recently been suggested to enhance the work
of adhesion, but the study in Ref. \cite{Kim} is for a periodic surface profile (sinus corrugation) and it is not cleat to what
extent the same effect contribute for randomly rough surface.

We have studied the influence of rapid local detachment processes (Fig. \ref{openingcrack}(b)) on the pull-off force
for the same system as above but for a curved substrate surface.
Thus we have curved the substrate surface so that it becomes a spherical cup with the
radius of curvature $R= 4410 {\rm \AA}$, but the surface roughness is the same as for the nominally flat surfaces
used above. We have measured the force on the block during approach and separation of the two solids, and used the JKR theory to determine
the effective interfacial energy $\gamma_{\rm eff}$. This approach includes the influence of local irreversible detachment
processes (Fig. \ref{openingcrack}(b)).
As above, the two solids were first approached to the point where the total force on the block vanished, and then separated.
We found that if the upper surface of the block moves with the velocity $v= 5 \ {\rm m/s}$ or less, the force on the block as a function
of the (average) block-substrate separation was independent of $v$ and the results presented below was obtained with $v=5 \ {\rm m/s}$.
We also tested that the damping which acted on the atoms (to remove elastic waves emitted from the interface)
was so small that it had no influence on the pull-off force. For the curved surface without surface roughness we obtained from
the maximum adhesion force using the JKR formula, for both the approach and the separation, the same interfacial energy
$\gamma_{\rm eff} =\Delta \gamma \approx 100 \ {\rm meV/\AA^2}$ as determined above for the flat
surfaces from the work done by the substrate on the bottom layer of the block.

\begin{figure}
\includegraphics[width=0.4\textwidth,angle=0]{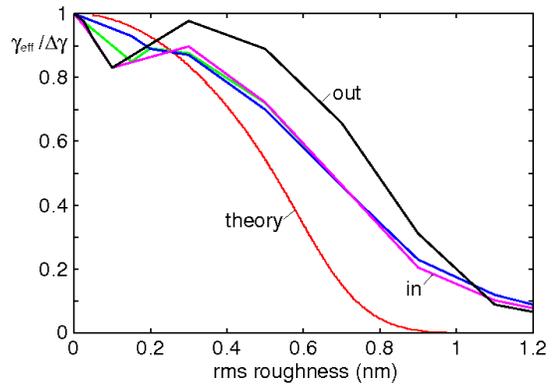}
\caption{\label{all}
The interfacial binding energy $\gamma_{\rm eff}$ (in units of the interfacial binding energy $\Delta \gamma = \gamma_1+\gamma_2-\gamma_{12}$
for flat surfaces) as a function of the root-mean-square roughness. The red, blue and green lines are from Fig. \ref{1rms.2gamma.exp.theory}.
The pink and black curves (denoted ``in'' and ``out'') are the effective interfacial energy during approach and during separation, respectively,
for the nominally curved (spherical cup) substrate (the curves was obtained from the maximum
adhesion force using the JKR theory).
The substrate surfaces are self-affine fractal (randomly rough) with the Hurst exponent $H=0.8$.
}
\end{figure}

In Fig. \ref{all} we show
the interfacial binding energy $\gamma_{\rm eff}$ (in units of the interfacial binding energy $\Delta \gamma = \gamma_1+\gamma_2-\gamma_{12}$
for flat surfaces) as a function of the root-mean-square roughness. The red, blue and green lines are from Fig. \ref{1rms.2gamma.exp.theory}.
The pink and black curves (denoted ``in'' and ``out'') are the effective interfacial energy during approach and during separation, respectively,
for the nominally curved (spherical cup) substrate (the curves was obtained from the maximum
adhesion force using the JKR theory).
Note that local irreversible detachment processes (in front of the crack tip) give a relative small increase in the pull-off force
for randomly rough surfaces, and have a negligible influence on the force on the block during contact formation (approach).

The rapid decrease in $\gamma_{\rm eff}$ observed in the MD simulations
for small surface roughness (${\rm rms} < 0.1 \ {\rm nm}$) may reflect atomistic binding
effects. That is, the minimum energy state for the flat surfaces correspond to the block atoms occupying hollow sites on the substrate.
For the rough surface other binding configurations may form, e.g., domain-wall structures which would
tend to increase the elastic energy stored at the interface and hence reduce $\gamma_{\rm eff}$. Such atomistic effects are, of course, not included in
the elastic continuum model.

\begin{figure}
\includegraphics[width=0.4\textwidth,angle=0]{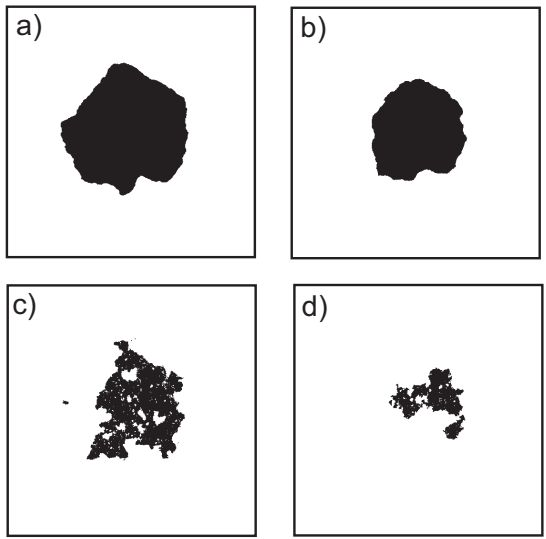}
\caption{\label{snap}
Snap shot pictures of the contact region during pull-off for contact between the curved
substrate with the rms roughness $3 \ {\rm \AA}$ [(a) and (b)] and $9 \ {\rm \AA}$ [(c) and (d)] and block surface.
Pictures (a) and (c) are for the point where the total force on the block vanish and (b) and (d)
for $0.14 \ {\rm ns}$ later where the upper surface of the block has moved $7 \ {\rm \AA}$ upwards.
}
\end{figure}

Fig. \ref{snap} shows snapshot pictures of the contact region during pull-off for contact between the curved
substrate with the rms roughness $3 \ {\rm \AA}$ [(a) and (b)] and $9 \ {\rm \AA}$ [(c) and (d)].
Fig. \ref{snap}(a) and (c) are for the point where the total force on the block vanish and (b) and (d)
for $0.14 \ {\rm ns}$ later where the upper surface of the block has moved $7 \ {\rm \AA}$ upwards.
Note that the contact region for  the rms roughness $3 \ {\rm \AA}$ is compact, i.e., $A/A_0 =1$
where $A_0$ is the nominal contact area, while for the rms roughness $9 \ {\rm \AA}$ the contact is
incomplete. This is consistent with the theory prediction [see Fig. \ref{1rms.2gamma.exp.theory}(b)] which shows complete contact
for the rms roughness $3 \ {\rm \AA}$ but only partial contact for the rms roughness $9 \ {\rm \AA}$.

Finally, we note that according to the continuum mechanics theory (see, e.g., \cite{P2}),
adhesive contact mechanics depends on the dimensionless parameter
$\theta = E^* h^2_{\rm rms} q_0/\Delta \gamma$ and we could have plotted
$\gamma_{\rm eff}/\Delta \gamma$ as a function of $\theta$ instead of
$h_{\rm rms}$ to emphasize the more general or universal nature of our results.
Of course, the result also depends on the detailed form of the surface roughness power
spectra used. The MD model is atomistic and depends on an additional dimension less
number, namely the Tabor number\cite{Tabor7} $\mu = R^{1/3}\Delta \gamma^{2/3} E^{-2/3}r_0^{-1}$,
where $r_0$ is an atomic bond distance. Estimation of the Tabor number for our
case shows that for the curved surface $\mu \approx 5$ so we are basically in the JKR-regime and
therefore the pull-off force is given to a good approximation by Eq. (6) rather
than the result of Derjaguin (valid for rigid solids) $F=2\pi R \gamma_{\rm eff}$, or some
intermediate value (e.g., DMT-theory)\cite{JG}. This is also confirmed by the
fact that for smooth surfaces the $\Delta \gamma$ deduced for the flat substrate surface is
the same as the $\Delta \gamma$ deduced from the JKR-formula for curved substrate surface (if the
Derjaguin theory would instead be valid, the $\Delta \gamma$ deduced using
the JKR theory should be a factor of $4/3$ times larger than deduced for the flat surface).

To summarize, we have presented a molecular dynamics (MD)
study of the adhesive contact between elastic solids with randomly
rough surfaces.
We have calculated the interfacial binding energy
and compared the results with the predictions of a recently developed contact mechanics model, which is based
on continuum mechanics. There is good general agreement between the MD-results and the continuum mechanics theory,
and the observed differences can be attributed to finite size effects and to the long range wall-wall interaction used in the
MD simulation, in contrast to the infinite system size and contact interaction assumed in the theory.
We have found that for randomly rough surfaces local irreversible detachment processes (in front of the crack tip) have a relative
small influence on the pull-off force and a negligible influence on the force on the block during contact formation (approach).


\begin{thebibliography}{999}

\bibitem{Bowden}
F.P. Bowden and D. Tabor, {\it Friction and Lubrication of Solids}
(Wiley, New York, 1956).

\bibitem{Johnson}
K.L. Johnson, {\it Contact Mechanics},
(Cambridge University Press, Cambridge, 1966).

\bibitem{BookP}
B.N.J. Persson,
{\it Sliding Friction: Physical Principles and Applications},
2nd edn. (Springer, Heidelberg, 2000).

\bibitem{Isra}
J.N. Israelachvili,
{\it Intermolecular and Surface Forces} (Academic, London (1995)).

\bibitem{JKRP}
K.L. Johnson, K. Kendall and A.D. Roberts,
Proc. R. Soc. Lond. A. {\bf 324}, 301 (1971)

\bibitem{Fuller}
K.N.G. Fuller and D. Tabor,
Proc. Roy. Soc. London A{\bf 345}, 327 (1975).

\bibitem{PerssonPRL}
A.G. Peressadko, N. Hosoda N and B.N.J. Persson, Phys. Rev. Lett. {\bf 95}, 124301 (2005).

\bibitem{Gao}
H. Gao and H. Yao,
PNAS{\bf 101} 7851 (2004)

\bibitem{Autumn}
K. Autumn and N. Gravish,
Phi. Tran. Roy. Sco. A.{\bf 366}, 1575 (2008)

\bibitem{Adhesion}
Y.P. Zhao, L.S. Wang and T.X. Yu,
J. Ad. Sci. Tech.{\bf 17}, 519 (2003)

\bibitem{all}
A.N. Gent, P.B. Lindley, Proc. R. Soc. Lond. A {\bf 249}, 195 (1958);
A. N. Gent, D. A. Tompkins, J. Applied Phys. {\bf 40}, 2520 (1969);
Lakrout, Sergot and Creton, J. Adhesion {\bf 69}, 307 (1999);
Gay and Leibler, Phys. Rev. Lett. {\bf 82}, 936 (1999);
Poivet et al. Europhys. Lett. {\bf 62}, 244 (2003); Eur. Phys. J. E{\bf 15},  97 (2004);
A. Zosel, Colloid and Polymer Sci. {\bf 263}, 541 (1985);
A. Zosel, J. Adhesion 30, 135 (1989) and Int. J. Adhesion \& Adhesives {\bf 18}, 265 (1998);
C. A. Dahlquist, Proc. Nottingham Conf. on Adhesion, 1966 Fundamental and Practice (MacLaren and Sons, Ltd. London);
C. A. Dahlquist, in Treatise on Adhesion and Adhesives, R. L. Patrick (ed.), Dekker, New York (1969), 2.

\bibitem{gorb}
B.N.J. Persson, A. Kovalev, M. Wasem, E. Gnecco and S.N. Gorb,
EPL {\bf 92}, 46001 (2010).

\bibitem{P4}
B.N.J. Persson, Phys. Rev. Lett. {\bf 99}, 125502 (2007)

\bibitem{P1}
B.N.J. Persson, J. Chem. Phys. {\bf 115}, 3840 (2001).

\bibitem{P2}
B.N.J. Persson, Eur. Phys. J E{\bf 8}, 385 (2002).

\bibitem{P3}
B.N.J. Persson, Surface Science Reports {\bf 61}, 201 (2006).

\bibitem{CarboneS}
G. Carbone, M. Scaraggi and U. Tartaglino, Eur. Phys. J E{\bf 30}, 65 (2009).

\bibitem{YP}
C. Yang and B.N.J. Persson, J. Phys. Condens. Matter {\bf 20}, 215214 (2008).

\bibitem{YTP}
C. Yang, U. Tartaglino and B.N.J. Persson,
Eur. Phys. J. E{\bf 19 }, 47 (2006)

\bibitem{Maugis1}
D. Maugis J. Coll. Interf. Sci. 150(1) 243-269, 1992

\bibitem{Carlos}
A. Almqqvist, C. Campana, N. Prodanov and B.N.J. Persson,
J.Mech. Phys. Solids {\bf 59}, 2355 (2011).

\bibitem{Gent}
A.N. Gent and J. Schultz, J. Adhesion {\bf 3}, 281 (1972).

\bibitem{Maugis}
D. Maugis and M. Barquins, J. Phys. D {\bf 11}, 1989 (1978).

\bibitem{Kim}
Q. Li and K-S Kim,
Acta Mechanica Solida Sinica {\bf 22}, 377 (2009).

\bibitem{Tabor7}
D. Tabor, J. Colloid Interface Sci. {\bf 58}, 2 (1977).

\bibitem{JG}
K.L. Johnson and J.A. Greenwood, J. Colloid Interface Sci. {\bf 192}, 326 (1997).

\end{thebibliography}
\end{document}